\documentclass [aps,prb,amssymb,amsmath,twocolumn, showpacs]{revtex4-1}
\usepackage[latin9]{inputenc}
\usepackage{amsmath}
\usepackage{amssymb}
\usepackage{graphicx}
\usepackage{verbatim}
\usepackage{bm}
\usepackage{color}

\usepackage{graphicx,epsfig,amsfonts,amssymb}
\usepackage{bm}
\usepackage{times}
\usepackage{verbatim}
\usepackage[colorlinks,citecolor=blue,linkcolor=blue,hyperindex]{hyperref}

\newcommand{\be}{\begin{eqnarray}}
\newcommand{\ee}{\end{eqnarray}}

\newcommand{\bs}{\begin{equation}\begin{split}}
\newcommand{\es}{\end{split}\end{equation}}

\begin{document}
\title{PH-Pfaffian order in a translationally and rotationally invariant system}

\author{Chen Sun$^{1,2}$, Ken K. W. Ma$^1$, and D. E. Feldman$^1$}
\affiliation{$^1$Brown Theoretical Physics Center and Department of Physics, Brown University, Providence, Rhode Island 02912, USA\\
$^2$School of Physics and Electronics, Hunan University, Changsha 410082, China}

\begin{abstract}
The PH-Pfaffian topological order has been proposed as a candidate order for the $\nu=5/2$ quantum Hall effect. The PH-Pfaffian liquid is known to be the ground state in several coupled wire and coupled stripe constructions.
No translationally and rotationally invariant models with the PH-Pfaffian ground state have been identified so far. By employing anyon condensation on top of a topological order, allowed in an isotropic system, we argue that the PH-Pfaffian order is possible in the presence of rotational and translational symmetries.
\end{abstract}

\maketitle

\section{Introduction}

The topological order on half-integer quantum Hall plateaus has been a subject of much debate \cite{MFreview,16-fold}. There has long been tension between experiment \cite{MFreview,Banerjee2018} and numerics \cite{Morf,Rezayi,Pakrouski}. It increasingly appears that
multiple topological orders are present in experimentally relevant systems. Indeed, numerical evidence exists for different topological orders on half-integer plateaus in GaAs \cite{Rezayi,Pakrouski,Luo} and graphene \cite{Wu}.
Some experiments \cite{Pan,Samkharadze} even hint at different topological orders at different magnetic fields on the $5/2$ plateau in GaAs.
Such behavior differs profoundly from the intuition that builds on the properties of the simplest and best-understood quantum Hall state at $\nu=1/3$, where the same Laughlin topological order \cite{Laughlin} is believed to be present in a broad range of materials and  parameters.

The difficulties with half-integer filling factors reflect a stronger role for composite-fermion (CF) interactions on half-integer plateaus than at most odd-denominator filling factors \cite{Jain_book}. Indeed, a great majority of odd-denominator states can be understood as integer quantum
Hall states of CFs. Such integer states are present even for non-interacting CFs, and their interaction does not affect qualitative features, such as possible topological orders.
In contrast to this picture, non-interacting CFs would not form an incompressible liquid at a half-integer filling \cite{Jain_book}. This agrees with the absence of the $1/2$ and $3/2$ plateaus in monolayer GaAs.
At the same time, experimental evidence exists for CFs on the quantized $5/2$ plateau \cite{Willett,Hossain}. This suggests that the $5/2$ plateau forms due to CF interactions.
The plateau can be explained by Cooper pairing of CFs \cite{RG2000}. The details of the topological order depend on the pairing channel: Different channels result in 8 possible Abelian and 8 possible non-Abelian orders \cite{16-fold}.

Which one or ones are present in experimentally relevant systems? Preponderance of numerical evidence \cite{MFreview,Rezayi,Pakrouski} points towards Pfaffian \cite{MG1991} and anti-Pfaffian \cite{APf_Levin2007,APf_Lee2007} liquids in translationally invariant systems.
Preponderance of experimental evidence  \cite{MFreview,Banerjee2018} suggests the PH-Pfaffian order \cite{APf_Lee2007,Son2015,Zucker2016,FCV,BNQ}
on the $5/2$ plateau in GaAs. A possible explanation of such discrepancy comes from disorder \cite{Zucker2016,Mross,Wang,Lian}, inevitable in any sample, but ignored in all numerical studies until a very recent paper \cite{Zhu}.
Weak disorder is not believed to affect topological order at $\nu=1/3$. Strong disorder destroys the $1/3$ plateau.
This behavior is the same as in the integer quantum Hall effect. At the same time,  disorder can change the pairing channel in a superconductor \cite{superconductivity}. This suggests that disorder may change the qualitative physics of the CF superconductor at $\nu=5/2$.
Recent theoretical work \cite{Mross,Wang,Lian} does predict a complicated phase diagram in the presence of disorder with several topologically ordered phases and a gapless thermal metal. Note that a random potential is not necessary for the stabilization of the PH-Pfaffian liquid.
Coupled wire constructions and a coupled stripe construction produce Hamiltonians with the PH-Pfaffian order in the ground state without any randomness \cite{16-fold,Kane-wire,Fuji-wire}.
The common feature of the disorder-based approach with those constructions consists in the absence of translational and rotational symmetry.

A possible lesson might be that the PH-Pfaffian order were impossible in uniform systems. Yet, it was suggested that it might be stabilized by sufficiently strong Landau level mixing (LLM) even in uniform systems \cite{Milovanovic1,Milovanovic2}.
If so, a translationally and rotationally invariant model should exhibit PH-Pfaffian order. In this paper we argue that the PH-Pfaffian order does emerge in isotropic systems as a result of anyon condensation \cite{Burnell,ERB,more-condensation} on top of another topological order.

The model system is multi-component. We argue that for appropriate microscopic interactions, the components may originate from different Landau levels. This makes our model different from constructions in which wave functions of various topological orders are localized in a single Landau level, and thus LLM is ignored.
This difference is consistent with recent numerical results \cite{gap0,gap}, which suggest that the PH-Pfaffian state loses its gap after projection into the lowest Landau level. It is also consistent with the symmetry-from-no-symmetry principle \cite{Zucker2016}, which postulates that a particle-hole symmetric topological order is only possible,
if the particle-hole symmetry is broken by LLM, disorder, or another mechanism or combination of mechanisms.

In what follows, we start with a review of the PH-Pfaffian topological order. We then observe that the edge structure of a PH-Pfaffian liquid can be obtained from a two-component system. One component is made of charged fermions and the other is made of neutral bosons. In the fourth section we argue that the two-component system
possesses the PH-Pfaffian order in the bulk. In the final section we propose a scenario how such two-component model might be realized in a purely electronic system.

\section{PH-Pfaffian order}

The anyons are labeled by their topological charge $t=1,\sigma,$~or $\psi$ and the electric charge $ne/4$, where $n$ is odd in the $\sigma$-sector and even otherwise. We will use the notation $(t,n)$. The fusion rules are

\begin{equation}
\label{dima-2020-1}
\psi\times\psi=1;~\sigma\times\psi=\sigma;~ \sigma\times\sigma=1+\psi,
\end{equation}
where $1$ stays for vacuum and $\psi$ is a Majorana fermion. The statistical phase, accumulated by an anyon of type $(t_1,n_1)$ while making a full counterclockwise circle around an anyon of type $(t_2,n_2)$
is

\begin{equation}
\label{dima-2020-2}
\phi=\phi_{\rm nA}(t_1,t_2,f)+\frac{\pi n_1n_2}{4}
\end{equation}
where the non-Abelian phase $\phi_{\rm nA}$ depends on the topological charges $t_1$ and $t_2$ and on the fusion channel $f$, Eq. (\ref{dima-2020-1}). The non-Abelian phase is trivial, $\phi_{\rm nA}=0$, if $t_1=1$, $t_2=1$, or $t_1=t_2=\psi$. For two $\sigma$-particles,
the non-Abelian phase depends on the fusion channel:

\begin{equation}
\label{dima-2020-3}
\phi_{\rm nA}(\sigma,\sigma,1)=\pi/4,~\phi_{\rm nA}(\sigma,\sigma,\psi)=-3\pi/4.
\end{equation}
Finally,

\begin{equation}
\label{dima-2020-4}
\phi_{\rm nA}(\sigma,\psi,\sigma)=\phi_{\rm nA}(\psi,\sigma,\sigma)=\pi.
 \end{equation}
The bulk statistics determines the edge Lagrangian density \cite{Zucker2016}:

\begin{eqnarray}
\label{dima-2020-5}
L=\frac{2}{4\pi}\partial_x\phi_c(\partial_t-v_c\partial_x)\phi_c
+i\psi(\partial_t+u\partial_x)\psi,
\end{eqnarray}
where $\psi$ is a Majorana fermion and the charge mode $\phi_c$ sets the charge density $e\partial_x\phi_c/2\pi$ on the edge. An edge excitation from the sector $(t,n)$ is created by the operator $t\exp(in\phi_c/2)$, where $t=1,\sigma,\psi$ acts in the neutral Majorana sector with $\sigma$ being the twist operator.
The electron operator is $\psi\exp(2i\phi_c)$. Both the thermal and electrical
conductances are one half of a quantum \cite{Son2015,Zucker2016}.
We include more details about the statistics in the PH-Pfaffian and related orders in Supplemental Material \cite{sup}.

\section{Model: View from the edge}

Our starting point is a two-component system. One component is a fractional quantum Hall (FQH) liquid in the anti-Pfaffian state \cite{APf_Levin2007,APf_Lee2007}. The other component is made of neutral bosons in the Laughlin state at the filling factor $\nu=1/4$.
Rotationally and translationally invariant models with those two orders in their ground states are known.

The anti-Pfaffian order is very similar to the PH-Pfaffian order. The classification of the excitations and their fusion rules are the same. Only a small difference exists in the braiding rules: the Abelian phase has the opposite sign compared to (\ref{dima-2020-2}):

\begin{equation}
\label{dima-2020-6}
\phi=\phi_{\rm nA}(t_1,t_2,f)-\frac{\pi n_1n_2}{4}.
\end{equation}
The edge theory differs from (\ref{dima-2020-5}) by the opposite propagation direction of $\phi_c$ and an additional charge integer mode $\phi_1$ with the charge density $e\partial_x\phi_1/2\pi$. The edge Lagrangian density

\begin{eqnarray}
\label{dima-2020-7}
L_{\rm aPf}=-\frac{2}{4\pi}\partial_x\phi_c(\partial_t+v_c\partial_x)\phi_c
+i\psi(\partial_t+u\partial_x)\psi\nonumber\\
+\frac{1}{4\pi}\partial_x\phi_1(\partial_t-v_1\partial_x)\phi_1+w\partial_x\phi_1\partial_x\phi_c.
\end{eqnarray}

Edge excitations are created by the same operators as in the PH-Pfaffian state. There are two electron operators: $\psi\exp(-2i\phi_c)$ and $\exp(i\phi_1)$. The operator $\exp(i\phi_n)=\exp(i[\phi_1+2\phi_c])$ creates a neutral fermionic excitation in the Majorana sector $\psi$.
The electrical conductance is half a quantum, as in the PH-Pfaffian state. The thermal conductance is $-1/2$ of a quantum \cite{APf_Levin2007,APf_Lee2007}.

It will be convenient to switch from the variables $\phi_c$ and $\phi_1$ to the neutral mode $\phi_n$ and the overall charge mode $\phi_\rho=\phi_1+\phi_c$. The Lagrangian density becomes

\begin{eqnarray}
\label{dima-2020-7b}
L_{\rm aPf}=\frac{2}{4\pi}\partial_x\phi_\rho(\partial_t-v_\rho\partial_x)\phi_\rho
+i\psi(\partial_t+u\partial_x)\psi\nonumber\\
-\frac{1}{4\pi}\partial_x\phi_n(\partial_t+v_n\partial_x)\phi_n+\tilde w\partial_x\phi_\rho\partial_x\phi_n.
\end{eqnarray}

The $\nu=1/4$ Laughlin state is Abelian \cite{Wen_book}. The phase accumulated by a fundamental anyon $\exp(ib)$ on a full counterclockwise circle around an identical anyon is $\pi/2$. The fusion of $n$ fundamental anyons yields a composite anyon $\exp(inb)$. Such anyon accumulates the phase $mn\pi/2$
on a full circle about an anyon of type $\exp(imb)$. As a consequence, $\exp(2ib)$ are fermions. $\exp(4ib)$ is topologically trivial. The edge theory of the Laughlin state assumes the form

\begin{equation}
\label{dima-2020-8}
L_{\rm B}=\frac{4}{4\pi}\partial_x b(\partial_t-v_b\partial_x)b.
\end{equation}
The electrical conductance of the neutral bosons is 0. The thermal conductance equals one quantum \cite{Kane_thermal}.

We now observe that the sums of the electric and thermal conductances of the bosonic liquid and the anti-Pfaffian liquid equal the electric and thermal conductances of the PH-Pfaffian liquid. This makes us expect that the PH-Pfaffian order should be present in a two-component system made of the anti-Pfaffian and bosonic Laughlin liquids.
We start with demonstrating that the edge structure of the PH-Pfaffian liquid can be obtained from such two-component model as illustrated in Fig. 1.

\begin{figure}[!htb]
\bigskip
\centering\scalebox{0.3}[0.3]{\includegraphics{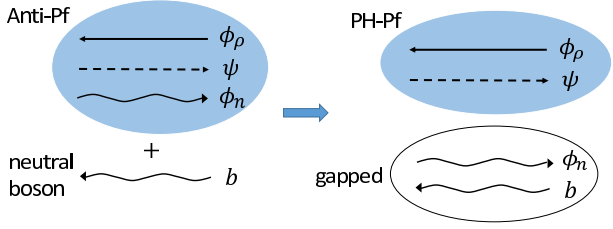}}
\caption{{Emergence of the PH-Pfaffian edge structure. Left panel: edge modes of two noninteracting layers with the anti-Pfaffian and Laughlin orders.
Right panel: the counterpropagating modes $\phi_n$ and $b$ are gapped out, and the remaining two modes exhibit  the edge structure of a PH-Pfaffian liquid.}}
\label{fig:edge-modes}
\end{figure}

We consider a two-component model with the following Lagrangian density on the edge:

\begin{equation}
\label{dima-2020-9}
L=L_{\rm aPf}+L_{\rm B}+u\partial_x\phi_n\partial_x b+U\cos(2\phi_n-4b).
\end{equation}
The cosine term is allowed in the action since it is topologically trivial and conserves the electric charge.
For simplicity, we assume \cite{footnote} that
$\tilde{w}=0$ in ${L}_{\rm aPf}$. The results do not change for a finite small $\tilde w$. The two counter-propagating modes $b$ and $\phi_n$ are gapped out if the cosine term is relevant in the renormalization group sense. After introducing a new field $\phi_b=-2b$, the contribution to the Lagrangian density that depends on $\phi_n$ and $\phi_b$ becomes
\begin{align}
\nonumber
{L}_{\rm n,b}
=
&\frac{1}{4\pi}
\left[\partial_x\phi_b\left(\partial_t-v_b\partial_x\right)\phi_b
-\partial_x\phi_n\left(\partial_t+v_n\partial_x\right)\phi_n\right]
\\
&-\frac{u}{2}\partial_x\phi_n \partial_x\phi_b
+U\cos{\left[2\left(\phi_n+\phi_b\right)\right]}.
\end{align}
 The stability of the edge requires
$|\pi u|\leq\sqrt{v_b v_n}$.

The Lagrangian density ${L}_{\rm n, b}$ can be diagonalized by the transformation~\cite{Wen_book}:
\begin{align}
\phi_b
&= \cosh{\theta}~\tilde{\phi_b}+\sinh{\theta}~\tilde{\phi}_n ,
\\
\phi_n
&= \sinh{\theta}~\tilde{\phi_b}+\cosh{\theta}~\tilde{\phi}_n,
\\
\tanh{2\theta}
&=-\frac{2\pi u}{v_b+v_n}.
\end{align}
In the new basis, the cosine term becomes
\begin{eqnarray}
{L}_{\rm tun}
=
U\cos{\left[2\left(\cosh{\theta}
+\sinh{\theta}\right)\left(\tilde{\phi}_n+\tilde{\phi_b}\right)\right]}.
\end{eqnarray}
Its scaling dimension can be deduced ~\cite{bosonization}:
\begin{align}
\Delta
=&~4\left(\cosh{2\theta}+\sinh{2\theta}\right)
=4\sqrt\frac{v_b+v_n-2\pi u}{v_b+v_n+2\pi u}.
\end{align}
When $\Delta<2$, ${L}_{\rm tun}$ is relevant and gaps out $\phi_b$ and $\phi_n$. This happens for
\begin{eqnarray}
\frac{3(v_b+v_n)}{10\pi} < u <\frac{ \sqrt{v_nv_b}}{\pi}.
\end{eqnarray}
The remaining two gapless modes $\phi_\rho$ and $\psi$ are described by the action identical to the PH-Pfaffian action (\ref{dima-2020-5}).

\section{Model: View from the bulk}

The action (\ref{dima-2020-9}) is the key to the bulk model. Indeed, $\cos(2\phi_n-4b)$ can be represented in the form $\hat B\hat B+ \hat B^\dagger \hat B^\dagger$, where $\hat B$ creates an excitation $B=(\psi,0)\times \exp(2ib)$. Such excitation is a product of two fermions and hence a boson.
The edge action thus suggests to consider the condensation of bosons $B$.
The condensation results in the confinement of many anyon types \cite{Burnell,ERB,more-condensation}. As we will see, the statistics of the remaining deconfined excitations is PH-Pfaffian. We argue that $B$ is condensable in two ways: using algebraic theory of anyons \cite{kitaev-16} in Supplemental Material \cite{sup}
and using the above edge construction in the end of this section.

Deconfined excitations braid trivially with $B$. Hence, the only non-trivial deconfined excitation of the Bose-liquid is $\exp(2ib)$. The deconfined excitations of the anti-Pfaffian liquid are $(\psi,2n)$ and $(1,2n)$. The attachment of any number of bosons $B$ does not change the superselection sector of an excitation. Thus, $\exp(2ib)$ and $\psi$ can be identified.

What about deconfined anyons that combine topological excitations of the Bose and anti-Pfaffian subsystems?
First, we can combine any number of deconfined excitations in the Bose and anti-Pfaffian sectors. This yields anyons of the types $(t,2n)\times\exp(2mib)$, where $t=1,\psi$. By attaching $(n-m)$ $B$-particles, any such anyon can be reduced to the standard type $(t',2n)\times\exp(2nib)$, where $t'=1,\psi$ is not necessarily the same as $t$.
In addition to products of deconfined excitations of the two subsystems, deconfined excitations exist in the $\sigma$ sector: $(\sigma,2n+1)\times\exp([2m+1]ib)$. Without loss of generality we can set $n=m$ since attaching $(n-m)$ bosons $B$ changes  $(\sigma,2n+1)\times\exp([2m+1]ib)$
into  $(\sigma,2n+1)\times\exp([2n+1]ib)$. Thus, all superselection sectors can be labelled as $(t,n)\times\exp(inb)$.

Neither of those sectors splits. Indeed, only non-Abelian anyons can split and only if the fusion of such an anyon with its antiparticle contains orthogonal copies of the vacuum of the condensed phase \cite{Burnell}. One sees that this does not happen in our problem.

We will now observe that all deconfined anyons can be identified with excitations of
a PH-Pfaffian liquid. We identify $(t,n)\times\exp(inb)$ with the $(t,n)$ anyon of the PH-Pfaffian order. All fusion rules are satisfied after such identification. The non-Abelian part of the braiding phase (\ref{dima-2020-2}) is also correct. The Abelian part of the mutual braiding phase of the anyons $(t_1,n_1)\times\exp(in_1b)$ and
$(t_2,n_2)\times\exp(in_2b)$
 is now the sum of the anti-Pfaffian contribution $-n_1n_2\pi/4$ and the Laughlin contribution $n_1n_2\pi/2$. This gives the correct PH-Pfaffian value.

 The above discussion assumes that $B$ is condensable. While this is plausible, can this be placed on a more rigorous footing? In addition to the discussion in Supplemental material \cite{sup}, we support this point with a coupled stripe construction (Fig. 2) in the spirit of Ref. \onlinecite{16-fold}.
 The bulk anti-Pfaffian and Laughlin orders can be obtained in a system of narrow stripes with anti-Pfaffian and Laughlin edge modes, in which counterpropagating modes of neighboring stripes gap each other. The charge modes are gapped out by the operators that tunnel charge 2$e$. Neutral modes are gapped out by operators
 that tunnel electrons and trivial bosons $\exp(4ib)$.
 We next add cosine terms $-A\cos(2b_L-\phi_{n,R})\cos(2b_R-\phi_{n,L})$ on each stripe, where the indices $L$ and $R$ show the right- and left-moving modes on the edges of the stripe. Such contribution creates trivial topological charge in each stripe. We also add interstripe tunneling between stripes $i$ and $i+1$
 of the form $-A'\cos(4b_{i+1,L}-2\phi_{n,i+1,R}-4b_{i,R}+
 2\phi_{n,i,L})$. We assume that the amplitudes $A$ and $A'$ are much greater than the amplitudes of any other tunneling terms. One sees that the system remains gapped in the bulk and the boson operator $\exp(2ib-i\phi_n)$ acquires a nonzero expectation value.
 This suggests that $B$ can condense in an anisotropic system. Since condensability is a topological property, the condensation of $B$ should also be possible in a rotationally and translationally invariant system.
 
 \begin{figure}[!htb]
\bigskip
\centering\scalebox{0.3}[0.3]{\includegraphics{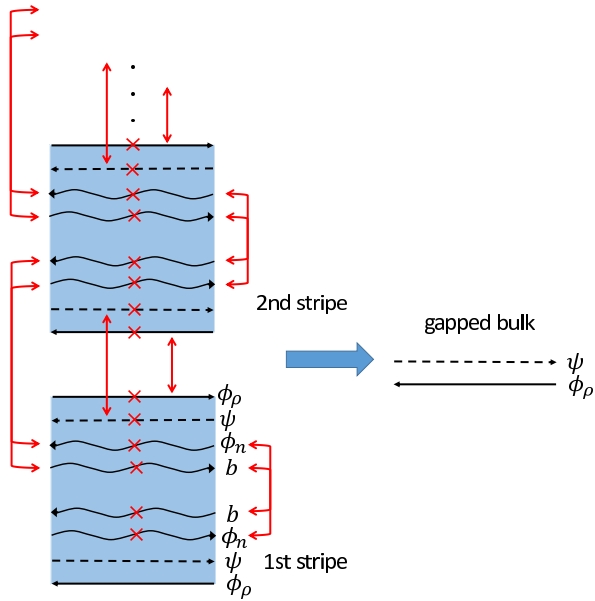}}
\caption{{A coupled-stripe construction of the PH-Pfaffian order. The red lines with arrows represent interactions which gap out the connected modes.}}
\label{fig2:edge-modes}
\end{figure}

\section{Conclusions}

The above model demonstrates that the PH-Pfaffian topological order can be obtained in a uniform system. All other known models \cite{16-fold,Mross,Wang,Lian,Kane-wire,Fuji-wire}  with that order break the translational and rotational symmetry either because of impurities or because the models consist of coupled wires or stripes.
Note that crystal structure implies that QHE systems are never exactly isotropic even in the absence of disorder, but this makes little difference at relevant electron densities.

Since our model combines fermions in the anti-Pfaffian state with neutral bosons,
its most natural realization would come from cold atoms. The model may seem
disconnected from the physics of the $5/2$ plateau in semiconductors, where only fermions are present. We propose a scenario that makes a connection with a purely fermionic system. We assume that electrons are present in four spin-resolved Landau levels. Electrons in one level exhibit the anti-Pfaffian order and form one of the two subsystems we need.
The electrical conductance of the anti-Pfaffian subsystem is one half of a quantum. The other three partially and fully filled Landau levels contribute two quanta to the electrical conductance, as necessary for the total conductance of $5/2$. One Landau level is fully filled. The sum of the filling factors of the other two is 1. Thus, the number of the holes in one of those Landau levels equals the number of the electrons in the other. We assume that all holes from one level combine with the electrons from the other level to form neutral bosons. The bosons form the Laughlin $\nu=1/4$ state, provided that their two-body
interaction favors the relative angular momentum $+4$, where the plus emphasizes that only one sign of the angular momentum along the $z$-axis is favored. An appropriate choice of the interaction between the bosons and the fermions in the anti-Pfaffian state yields the desired model system \cite{sup}.

The boson interaction breaks the time-reversal symmetry. This property is not shared by the Coulomb interaction in realistic samples. The time-reversal symmetry is broken instead by the external magnetic field to which neutral bosons do not minimally couple. Even if the interaction with the magnetic field is the only  contribution to the microscopic Hamiltonian that breaks the time-reversal symmetry, it is possible that additional symmetry-breaking interactions are generated in the effective low-energy Hamiltonian. Of course, it may well be that this does not happen for realistic Coulomb interactions. The point of our model is to show that the PH-Pfaffian order is possible without breaking the translational symmetry. More research is needed
to understand if the PH-Pfaffian order could be stabilized in realistic semiconductor heterostructures in the absence of random impurities.

\section*{Acknowledgements} We acknowledge a useful discussion with M. V. Milovanovi\'c. This work was supported by the J. Michael Kosterlitz Postdoctoral Fellowship at Brown University (C. S.), the Galkin Fellowship (K. K. W. M.) and by the NSF grant No. DMR-1902356.

\appendix

\section{Supplemental Material for ``PH-Pfaffian order in a translationally and rotationally invariant system''}

All sections of this Supplemental Material rely on algebraic theory of anyons \cite{kitaev-16}.

\subsection{Description of topological orders}

The purpose of this section is to summarize topological data of the Pfaffian, anti-Pfaffian, PH-Pfaffian, and 
$\nu=1/4$ Laughlin topological orders. The last order is known as $Z_4^{1/2}$ in the literature on algebraic theory of anyons. Nothing in this section is new. We build on tables from Ref. \onlinecite{bonderson-PhD}.

We start with the fusion rules for the Pfaffian state. They are summarized in Table~\ref{tab:fusion-Pf}. Up to a change of notation, the table reproduces Table A.1.24 from Ref. \onlinecite{bonderson-PhD}. Note that the table includes 12 anyon types, in an apparent contradiction with the standard physical picture of 6 anyons \cite{6types}. The reason is that the Pfaffian order is not modular: 
Electrons are not bosons but braid trivially with all excitations. This point has little physical consequence but allows a more rigorous discussion. 

\small{
\begin{widetext}
\center
\begin{table} [htb]
\scriptsize
\renewcommand{\arraystretch}{1.2}
\begin{tabular} {| c | c | c | c | c | c | c | c | c | c | c | c | c |}
\hline 
 ~$\times$~ & ~$(1,0)$~ & ~$(1,2)$~  & ~$(1,4)$~ & ~$(1,6)$~ & ~$(\sigma, 1)$~ & ~$(\sigma, 3)$~ & ~$(\sigma, 5)$~ & ~$(\sigma, 7)$~ & ~$(\psi, 0)$~ & ~$(\psi, 2)$~ & ~$(\psi, 4)$~ & ~$(\psi, 6)$~ \\ \hline
~$(1,0)$~ & ~$(1,0)$~ & ~$(1,2)$~  & ~$(1,4)$~ & ~$(1,6)$~ & ~$(\sigma, 1)$~ & ~$(\sigma, 3)$~ & ~$(\sigma, 5)$~ & ~$(\sigma, 7)$~ & ~$(\psi, 0)$~ & ~$(\psi, 2)$~ & ~$(\psi, 4)$~ & ~$(\psi, 6)$~
\\ \hline
~$(1,2)$~ & ~$(1,2)$~ & ~$(1,4)$~  & ~$(1,6)$~ & ~$(1,0)$~ & ~$(\sigma, 3)$~ & ~$(\sigma, 5)$~ & ~$(\sigma, 7)$~ & ~$(\sigma, 1)$~ & ~$(\psi, 2)$~ & ~$(\psi, 4)$~ & ~$(\psi, 6)$~ & ~$(\psi, 0)$~ 
\\ \hline
~$(1,4)$~ & ~$(1,4)$~ & ~$(1,6)$~  & ~$(1,0)$~ & ~$(1,2)$~ & ~$(\sigma, 5)$~ & ~$(\sigma, 7)$~ & ~$(\sigma, 1)$~ & ~$(\sigma, 3)$~ & ~$(\psi, 4)$~ & ~$(\psi, 6)$~ & ~$(\psi, 0)$~ & ~$(\psi, 2)$~
\\ \hline
~$(1,6)$~ & ~$(1,6)$~  & ~$(1,0)$~  & ~$(1,2)$~ & ~$(1,4)$~ & ~$(\sigma, 7)$~ & ~$(\sigma, 1)$~ & ~$(\sigma, 3)$~ & ~$(\sigma, 5)$~ & ~$(\psi, 6)$~ & ~$(\psi, 0)$~ & ~$(\psi, 2)$~ & ~$(\psi, 4)$~
\\ \hline
~$(\sigma,1)$~ & ~$(\sigma,1)$~  & ~$(\sigma,3)$~  & ~$(\sigma,5)$~ & ~$(\sigma,7)$~ & ~$(1, 2)+(\psi,2)$~ & ~$(1, 4)+(\psi,4)$~ & ~$(1, 6)+(\psi,6)$~ & ~$(1, 0)+(\psi,0)$~ & ~$(\sigma, 1)$~ & ~$(\sigma, 3)$~ & ~$(\sigma, 5)$~ & ~$(\sigma, 7)$~ 
\\ \hline
 ~$(\sigma,3)$~  & ~$(\sigma,3)$~  & ~$(\sigma,5)$~  & ~$(\sigma,7)$~ & ~$(\sigma,1)$~ & ~$(1, 4)+(\psi,4)$~ & ~$(1, 6)+(\psi,6)$~ & ~$(1, 0)+(\psi,0)$~ & ~$(1, 2)+(\psi,2)$~ & ~$(\sigma,3)$~ & ~$(\sigma, 5)$~ & ~$(\sigma, 7)$~ & ~$(\sigma, 1)$~ 
\\ \hline
~$(\sigma,5)$~ & ~$(\sigma,5)$~ & ~$(\sigma,7)$~  & ~$(\sigma,1)$~ & ~$(\sigma,3)$~ & ~$(1, 6)+(\psi,6)$~ & ~$(1, 0)+(\psi,0)$~ & ~$(1, 2)+(\psi,2)$~ & ~$(1, 4)+(\psi,4)$~ & ~$(\sigma,5)$~ & ~$(\sigma, 7)$~ & ~$(\sigma, 1)$~ & ~$(\sigma, 3)$~ 
\\ \hline
~$(\sigma,7)$~ & ~$(\sigma,7)$~ & ~$(\sigma,1)$~  & ~$(\sigma,3)$~ & ~$(\sigma,5)$~ & ~$(1, 0)+(\psi,0)$~ & ~$(1, 2)+(\psi,2)$~ & ~$(1, 4)+(\psi,4)$~ & ~$(1, 6)+(\psi,6)$~ & ~$(\sigma,7)$~ & ~$(\sigma, 1)$~ & ~$(\sigma, 3)$~ & ~$(\sigma, 5)$~ 
\\ \hline
~$(\psi,0)$~ & ~$(\psi,0)$~ & ~$(\psi,2)$~  & ~$(\psi,4)$~ & ~$(\psi,6)$~ & ~$(\sigma,1)$~ & ~$(\sigma, 3)$~ & ~$(\sigma, 5)$~ & ~$(\sigma, 7)$~ & ~$(1,0)$~ & ~$(1, 2)$~ & ~$(1,4)$~ & ~$(1,6)$~ 
\\ \hline
~$(\psi,2)$~ & ~$(\psi,2)$~ & ~$(\psi,4)$~  & ~$(\psi,6)$~ & ~$(\psi,0)$~ & ~$(\sigma,3)$~ & ~$(\sigma, 5)$~ & ~$(\sigma, 7)$~ & ~$(\sigma, 1)$~ & ~$(1,2)$~ & ~$(1, 4)$~ & ~$(1,6)$~ & ~$(1,0)$~ 
\\ \hline
~$(\psi,4)$~ & ~$(\psi,4)$~ & ~$(\psi,6)$~  & ~$(\psi,0)$~ & ~$(\psi,2)$~ & ~$(\sigma,5)$~ & ~$(\sigma, 7)$~ & ~$(\sigma, 1)$~ & ~$(\sigma, 3)$~ & ~$(1,4)$~ & ~$(1, 6)$~ & ~$(1,0)$~ & ~$(1,2)$~ 
\\ \hline
~$(\psi,6)$~ & ~$(\psi,6)$~ & ~$(\psi,0)$~  & ~$(\psi,2)$~ & ~$(\psi,4)$~ & ~$(\sigma,7)$~ & ~$(\sigma, 1)$~ & ~$(\sigma, 3)$~ & ~$(\sigma, 5)$~ & ~$(1,6)$~ & ~$(1, 0)$~ & ~$(1,2)$~ & ~$(1,4)$~ 
\\ \hline
\end{tabular}
\caption{Fusion algebra of anyons in the Pfaffian state.}
\label{tab:fusion-Pf}
\end{table}
\end{widetext}
}

\normalsize

All excitations in the above table are products of anyons from the Ising topological order and the $Z_8^{1/2}$ Abelian theory, with the numbers $n$ in the notation $(t,n)$ representing superselection sectors of the $Z_8^{1/2}$ order. The topological data of the $Z_N^{1/2}$ theory with an arbitrary even $N$ are shown in Table~\ref{tab:data-ZN}, which is based on the second table of section 5.1 in Ref. \onlinecite{bonderson-PhD}. The topological data of the Ising theory are summarized in Table~\ref{tab:data-Ising}, which is based on Table 5.6 of Ref. \onlinecite{bonderson-PhD}. All data of the Pfaffian order can be obtained by multiplying appropriate entries from Tables~\ref{tab:data-ZN} and~\ref{tab:data-Ising} for $N=8$. 

\begin{table*} [tb] 
\centering
\renewcommand{\arraystretch}{2.0}
\begin{tabular}{| c | c | } 
\hline
\multicolumn{2}{| c |}{~\text{Fusion rule: }$[m]_N\times [n]_N=[m+n]_N$~ }
\\ \hline
\quad
$\begin{aligned}
\left[F^{m,n,p}_{m+n+p}\right]_{m+n, n+p}
=e^{i\pi m\left(\left[n\right]_N+\left[p\right]_N-\left[n+p\right]_N\right)/N}
\end{aligned}$
\quad\quad & \quad
$R^{m,n}_{m+n}=e^{\pi i [m]_N [n]_N/N}$~ 
\\ \hline
~$S_{m,n}=\frac{1}{\sqrt{N}}e^{2i\pi mn/N}$~ & 
~$M_{m,n}=e^{2i\pi mn/N}$~
\\ \hline
~$d_n$=1~ & ~$\theta_n=e^{i\pi[n]_N^2/N}$~
\\ \hline
~$c_N=1$~ & ~$\mathcal{D}=\sqrt{N}$ \quad\quad
\\ \hline
\end{tabular}
\caption{Topological data of the $Z_N^{1/2}$ anyon model. Here, $m,n,p\in\left\{0,1,\cdots, N-1\right\}$. The symbol $[n]_N$ denotes the least residue of $n~\text{mod } N$.}
\label{tab:data-ZN}
\end{table*}

Tables from Ref.~\cite{bonderson-PhD} do not explicitly contain the Frobenius-Schur indicators of anyons. These invariants are defined for particles which are their own antiparticles. They are given in Table 1 of Ref. \cite{kitaev-16} for the Ising order and are equal to 1 for all excitations. They are also 1 for all excitations of the $Z_4^{1/2}$ and $Z_8^{1/2}$ orders, as can be seen, e.g., from Proposition 2.6 in Ref. \onlinecite{RSW}.

\begin{table*} [tb]
\centering
\renewcommand{\arraystretch}{2.0}
\begin{tabular}{| c | c | } 
\hline
\multicolumn{2}{| c |}{~\text{Fusion rules: } $1\times t=t$, \quad $\sigma\times\sigma=1+\psi$, \quad $\sigma\times\psi=\sigma$, \quad $\psi\times\psi=1$~ }
\\ \hline
\quad
$\begin{aligned} 
\\
\left[F^{\sigma\sigma\sigma}_{\sigma}\right]_{rs}
&=\begin{pmatrix}
1/\sqrt{2} & 1/\sqrt{2} \\
1/\sqrt{2} & -1/\sqrt{2}
\end{pmatrix}_{rs}
\\[3mm]
\left[F^{\sigma\psi\sigma}_{\psi}\right]_{\sigma\sigma}
&=\left[F^{\psi\sigma\psi}_{\sigma}\right]_{\sigma\sigma}
=-1 \\[8pt]
\end{aligned}$ 
\quad\quad &  \quad
$\begin{aligned}
R^{\sigma\sigma}_1&=e^{-i\pi/8}, \quad R^{\sigma\sigma}_{\psi}=e^{3i\pi/8}
\\[3mm]
R^{\sigma\psi}_\sigma&=R^{\psi\sigma}_\sigma=e^{-i\pi/2}, \quad
R^{\psi\psi}_1=-1
\end{aligned}
$ \quad\quad
\\  \hline
$\begin{aligned}
\\
S=\frac{1}{2}
\begin{pmatrix}
1 & \sqrt{2} & 1 \\
\sqrt{2} & 0 & -\sqrt{2} \\
1 & -\sqrt{2} & 1
\end{pmatrix}
\\[8pt]
\end{aligned}$
& 
$M=
\begin{pmatrix}
1 & 1 & 1 \\
1 & 0 & -1 \\
1 & -1 & 1
\end{pmatrix}
$
\\ \hline
$d_1=1$, $d_\sigma=\sqrt{2}$, $d_\psi=1$
& 
$\theta_1=1$, $\theta_\sigma=e^{i\pi/8}$, $\theta_\psi=-1$
\\ \hline
$c=1/2$ & $\mathcal{D}=2$
\\ \hline
\end{tabular}
\caption{Topological data of the Ising anyon model. The symbols $t\in\left\{1,\sigma,\psi\right\}$, and $r,s\in\left\{1,\psi\right\}$.}
\label{tab:data-Ising}
\end{table*}

The anti-Pfaffian order has the same set of excitations and the same fusion rules as the Pfaffian order. The data of the anti-Pfaffian order are obtained from the data of the Pfaffian order by the complex conjugation of Tables \ref{tab:data-ZN} and \ref{tab:data-Ising}.

We next proceed to the PH-Pfaffian order. It is a very close relative of the Pfaffian order. The fusion rules, Table~\ref{tab:fusion-Pf}, do not change. The only change in the rest of the data consists in the complex conjugation of Table~\ref{tab:data-Ising}.

Finally, the $Z_4^{1/2}$ order is fully described by Table~\ref{tab:data-ZN} at $N=4$ with $n$ and $m$ corresponding to the excitations $\exp(inb)$ and $\exp(imb)$.

\subsection{Condensability of the boson $\psi\exp(2ib)$}

According to Ref.~\cite{ERB}, a necessary condition of condensability consists in the trivial topological spin and the Frobenius-Schur indicator of 1. As seen from the previous section, the particle $B=\psi\exp(2ib)$ satisfies these conditions. The putative topological order on top of the Bose condensate of $B$ consists of those anyons that braid trivially with $B$. They are listed in the main text. Neither of them splits, as discussed in the main text. The goal of this section is to show that $B$ is in fact condensable, and the above putative topological order is hence well-defined.

A general approach to checking condensability~\cite{ERB} is ill-suited to practical use, and we apply a different {\it ad hoc} method. It reflects particular simplicity of our problem: $B$ is its own anti-particle, it has a unique fusion channel with each anyon type, and most importantly, the set $U$ of anyons that braid trivially with $B$ has a simple structure. As discussed in the main text, a subset $A$ of those anyons is closed with respect to braiding and fusion and each anyon of $U$ is either from $A$ or can be represented as  a fusion $a\times B$ of $B$ with $a\in A$. The set $A$ consists of anyons of the form $(t,n)\times\exp(inb)$, where $n$ is even for $t=1,\psi$ and $n$ is odd for $t=\sigma$. We first show that the set $A$ exhibits the PH-Pfaffian order. To prove that $B$ is condensable we then verify that anyons $a\times B$ exhibit the same fusion and braiding properties as $a$.

We rely on the bulk-edge correspondence \cite{RMP-bulk-edge}: wave functions of anyons can be expressed as conformal blocks of CFTs. In general, this correspondence is a conjecture, but it is well established for the Pfaffian and Laughlin orders. We will not need anything beyond those orders.

The Ising sector is described by the chiral CFT of a single Majorana fermion,
\begin{equation}
\label{1}
L=i\psi(\partial_t+\partial_x)\psi.
\end{equation}
The excitations are vacuum $1$, fermion $\psi$, and the Ising spin or twist field $\sigma$.
The Abelian $Z_N^{1/2}$ sectors are described by free-boson chiral CFTs with the Lagrangians of the form
\begin{equation}
\label{2}
L=-\frac{N}{4\pi}\partial_x\phi(\pm\partial_t+\partial_x)\phi,
\end{equation}
where the sign in the brackets determines the chirality, that is, the propagation direction of edge excitations in the system with the appropriate bulk order.  The excitation operators are $\exp(in\phi)$. The main text denotes $\phi=b$ at $N=4$ and $\phi=\phi_c/2$ at $N=8$. 
A subtlety involves the bulk-edge correspondence for Abelian orders, such as $Z_8^{1/2}$. Multiple operators $\exp(i\phi[n+8m])$ represent the same superselection sector. Thus, multiple wave functions need to be identified. Physically, they differ by  the local electric charge $e(n+8m)/4$. From the experimental point of view, this is arguably a greater distinction than subtle differences between superselection sectors.

We start with anyons from the set $A$. The wave function is the product of a conformal block of the operators $1,\sigma$, and $\psi$ in the theory (\ref{1}) and a conformal block of operators 
$\exp(in_k\phi_c/2+in_k b)$ in the combination of the theories (\ref{2}) at $N=8$ and $N=4$ with opposite chiralities. The latter conformal block is a polynomial of the complex coordinates $z$ and $\bar z$: 
\begin{equation}
\label{3}
C_A=\Pi_{i>j}(z_i-z_j)^{n_i n_j/8}|z_i-z_j|^{n_i n_j/4}.
\end{equation}
As usual, we omit the exponential factors 
$\exp(-{\rm const}\sum |z_i|^2)$ in the wave functions.
The absolute value factor in Eq. (\ref{3}) does not affect topological properties and can also be ignored. Then the structure of the wave functions makes it obvious that the set $A$ exhibits the PH-Pfaffian order. Indeed, the only difference from a Pfaffian conformal block consists in the complex conjugation of the Ising part. 

We now turn to the full theory that includes both the set $A$ and the set $A_B$ of anyons $a\times B$, $a\in A$. 

\begin{figure}[htb]
\includegraphics[width=3.2in]{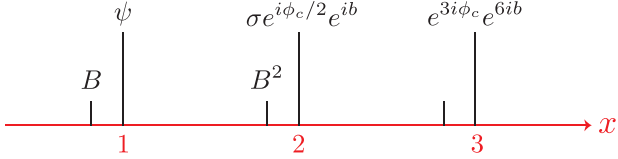}
\caption{An arrangement of operators in a conformal block.}
\label{fig:block}
\end{figure}

In the algebraic theory of anyons, it is conventional to order all anyons along a line from left to right. We will assume that anyons are placed at points with integer coordinates $l=1, 2, \dots$. More precisely, an anyon from the superselection sector $a\times B^j$, $j\in Z$, where $a\in A$, 
is represented by two operators: an operator $a=t\exp(i n\phi_c/2)\exp(i n b)$, with $t=1,\psi$ for an even $n$ and $t=\sigma$ for an odd $n$, at point $l$ and the operator $B^j=\psi^j\exp(2i j b)$ at point $l-\delta$, where $\delta\ll 1$. Many of those operators describe the same superselection sector. For example $B^2$ is equivalent to $1$. It will still be convenient to consider all corresponding conformal blocks and identify some of them. An example of an arrangement of operators is shown in Fig.~\ref{fig:block}. It corresponds to the conformal block 
$\langle B(x=1-\delta)\psi(x=1) B^2(x=2-\delta) [\sigma e^{i\phi_c/2}e^{ib}](x=2)
[e^{3i\phi_c}e^{6ib}](x=3)\rangle$, where additional operators in far away points are implied to make sure that the average is nonzero.

\begin{figure}[htb]
\includegraphics[width=3.0in]{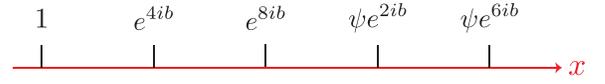}
\caption{An example of canonical ordering of operators $b_n$.}
\label{fig:ordering}
\end{figure}

We build the wave function $\tilde\Psi=\langle\Pi_l B_l(x=l-\delta)a_l(x=l)\rangle$ with $a_l\in A$, $B_l=B^j$ in two steps. We first consider the wave function $\Psi_0=\langle\Pi_la_l(x=l)\rangle \Psi^B_0$, where $\Psi^B_0$ does not depend on the anyon configuration under consideration and represents the wave function of ``boson storage'' far on the left from $x=0$. We assume that the total topological charge of boson storage is 1, that is, $\Psi^B_0=\langle\Pi_n b_n(x=-10^{10}-n)\rangle$ with $b_n=\psi^{m_n}\exp(2im_nb)$,  $\sum_n m_n=0$. This ensures that a conformal block built from the operators $b_n$ and $a_l$ factorizes into the product of a conformal block of $a_l$ and the correlation function of $b_l$. On the second step, we obtain a conformal block $\Psi$ from $\Psi_0$ by moving some of the operators $b_n$ to points with positive coordinates $x=l-\delta$. Their trajectories and the order in which they move do not matter since $B$ braids trivially with all excitations. It is essential that the excitations be ordered canonically in the storage before and after step 2. For example, we can choose the following ordering: operators $b_n=\psi\exp(ipb)$ are always on the right of all operators $\exp(iqb)$. The operator $\psi\exp(ik_1b)$ is on the right of $\psi\exp(ik_2b)$ if $k_2<k_1$. The operator $\exp(ik_1b)$ is on the right of  $\exp(ik_2b)$ if $k_2<k_1\ne 0$. Operators $b_n=1$ are placed on the left of all other operators. The canonical ordering is illustrated in Fig.~\ref{fig:ordering}. The canonical ordering is needed because $\Psi$ is not the same as $\tilde\Psi$ above. The ordering allows a natural identification of $\Psi$ and $\tilde\Psi$. Of course, we also identify an infinite number of conformal blocks that differ by factors like $\exp(4ib)$.

\begin{figure}[htb]
\includegraphics[width=3.3in]{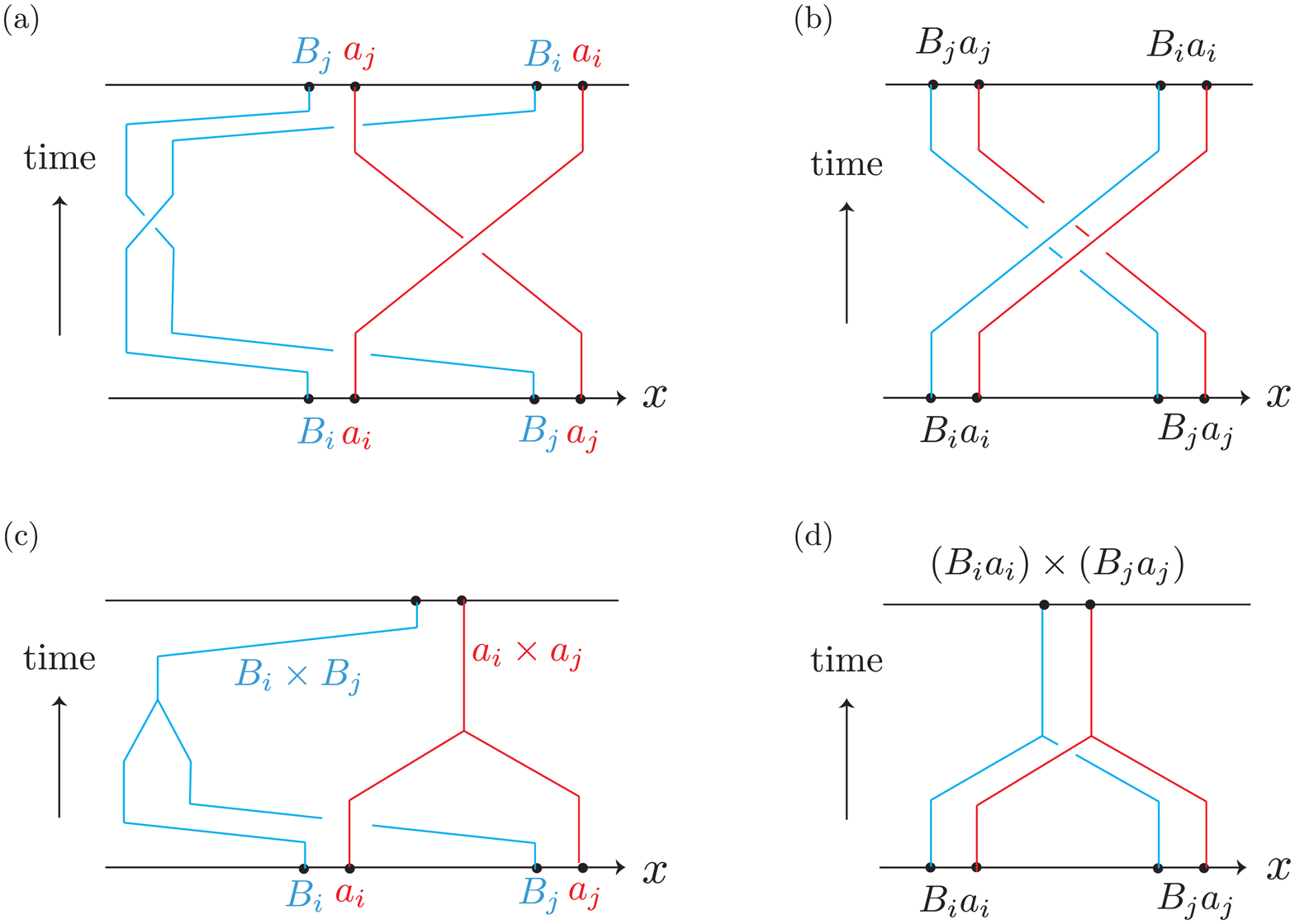}
\caption{Braiding and fusion of anyons $a\times B$. The red lines (blue lines) denote the trajectories of type-$a$ anyons (type-$B$ anyons).}
\label{fig:anyon}
\end{figure}

We perform braiding and fusion operations in three steps. First we move all $B^j$ from points $l-\delta$ far to the left. Next we braid or fuse operators $a_l$ and operators $B^j$ separately, and then we move the results of the fusion or braiding of $B^j$ to the appropriate points $l-\delta$. Figs.~\ref{fig:anyon}a and~\ref{fig:anyon}c illustrate this procedure for the braiding of two anyons and for fusing two anyons. Crucially, braiding and fusion act trivially on the set of $B^j$ ($F$ and $R$ symbols reduce to 1). Hence, diagrams~\ref{fig:anyon}a and~\ref{fig:anyon}c correspond to the same $F$ and $R$ symbols as in the absence of any $B^j$ factors, that is, the same $F$ and $R$ symbols as in the PH-Pfaffian theory. Diagrams~~\ref{fig:anyon}b and~~\ref{fig:anyon}d illustrate braiding and fusion without splitting $B^j$ away from $a_l$. They correspond to the conventional definition of braiding and fusion. The diagrams are topologically equivalent to~~\ref{fig:anyon}a and~\ref{fig:anyon}c. Hence, they generate the same linear transformations in the space of conformal blocks. This proves that the statistics is indeed PH-Pfaffian.

 One may raise an objection about splitting anyons. Indeed, the above approach assumes that the anyon $B^j\times a_l$ can only be split into $B^k \times a_n$ and $B^{j-k} \times a_m$ as $B^j\rightarrow B^k\times B^{j-k}$, $a_l\rightarrow a_m\times a_n$. What if we split $a_l$ and $B_j$ in a different way and then rearrange the parts to produce
 $B^k(x=1-\delta)a_n(x=1)B^{j-k}(x=2-\delta)a_m(x=2)$? It turns out that this generates the same linear transformation up to a constant factor, {\it i.e.}, the dimension of the corresponding splitting space is 1. This follows from two observations. First, all fusion multiplicities are 1 in the theory under consideration. Second, $B$ has a unique fusion channel with any anyon.

\subsection{Bose condensate}

In this section we propose a possible Hamiltonian with a Bose condensate of $B$ in its ground state. 

Our starting point is a two-layer system with the Laughlin $\nu=1/4$ order and the anti-Pfaffian order in its two layers. Hamiltonians are known for these types of orders \cite{Greiter-book} with the property that all particles are confined in the same Landau level. All other Landau levels will be ignored below. Indeed, one can add contributions to the Hamiltonian that will make them arbitrarily high in energy. The effective Hamiltonian in the lowest Landau level will be expressed in terms of the positions of anyonic excitations. For well-separated anyons, the energy is given by the sum of the appropriate energy gaps. For nearby anyons at the distance of the order of the magnetic length $l_B$, two difficulties are present. First, the energy is affected by their interaction. Second, the standard wave functions \cite{RMP-bulk-edge} of anyons in nearby points are not orthogonal. To avoid those complications, we introduce a repulsive interaction $V({\bf r}_1-{\bf r}_2)$ for anyons at points ${\bf r}_{1,2}$ such that $l<|{\bf r}_1-{\bf r}_2|<L$, where $L\gg l\gg l_B$. The interaction must change slowly on the scale of $l_B$ and become strong at  $|{\bf r}_1-{\bf r}_2| \sim l$. This ensures that anyons remain separated by the scale $\gg l$. The interaction is zero at $|{\bf r}_1-{\bf r}_2|<l$. This does not mean that multiple anyons are allowed in close quarters on the scale 
$l$. Instead, we treat several anyons within such region as a single anyon which is the outcome of their fusion.  It will be convenient to assume that only one state is possible in each superselection sector for a system of size $l$. The rest of the states can be pushed up in energy by appropriate local contributions to the Hamiltonian.

Next, we introduce a pair creation operator $\hat p$ for the boson $B$.  By definition, $\hat p({\bf r}_1,{\bf r_2})$ creates two bosons at points ${\bf r}_{1,2}$. This is a legitimate operator since the combined topological charge of the two bosons is trivial.
In algebraic theory of anyons, the diagram corresponding to the creation of two bosons is a line from ${\bf r}_1$ to ${\bf r}_2$. The line describes the trajectories of a particle and an antiparticle, created at the same point. We can assume that the trajectories of the two bosons are straight, if the straight line from ${\bf r}_1$ to ${\bf r}_2$ does not go within the distance $L$ from other anyons. Otherwise, the trajectories should bend to avoid coming within the radius of action of $V$.
The prescription for bending must respect the rotational symmetry. We do not dwell on its details since we will consider a situation with a low concentration of anyons in the system. For the same reason we will ignore any complications from the possibility that $|{\bf r}_2-{\bf r}_2|<L$. We next introduce the superposition operator
$\hat P({\bf x}_1)=\int_{|{\bf x}_1-{\bf x}_2|<D}  d^2 x_2 \hat p({\bf x}_1,{\bf x}_2)$, where $D\gg L$. To ensure translational invariance, the operator $\hat P({\bf y})$ should be obtained from $\hat P({\bf 0})$ with an appropriate magnetic translation.

The proposed Hamiltonian expresses in terms of $\hat P$:
\begin{align} \label{4}
\nonumber
H
=~&\frac{\tilde U}{2} 
\int d^2 x (\hat P({\bf x})+\hat P^\dagger({\bf x})) 
+\frac{\tilde W}{S} \int d^2 x \hat P^\dagger({\bf x}) \hat P({\bf x})
\\
&+\frac{\epsilon_0 N_B}{2},
\end{align}
where $\tilde U$ and $\tilde W$ are constants, $S=\frac{\pi D^2}{4}$, $N_B$ is the total number of the bosons, and $\epsilon_0$ is their energy gap. One can also add a gradient contribution to obtain a more realistic effective Hamiltonian, but we will focus on the simplest expression (\ref{4}).
We argue that the ground state of the above Hamiltonian is a Bose condensate of $B$ for sufficiently large $\tilde U$ and $\tilde W$. For this purpose we rely on mean-field theory. Thus, we set $D$ equal to the system size. The mean-field Hamiltonian
\begin{equation}
\label{5}
H_{\rm MF}=\frac{\tilde U}{2}  (\hat P+\hat P^\dagger) +\frac{\tilde W}{S} \hat P^\dagger \hat P+\frac{\epsilon_0 N_B}{2},
\end{equation}
where $\hat P=\int d^2 x_1 d^2 x_2 \hat p({\bf x}_1, {\bf x}_2)$ and the integration extends over the whole system area $S$.

\begin{figure} [htb]
\includegraphics[width=3.25in]{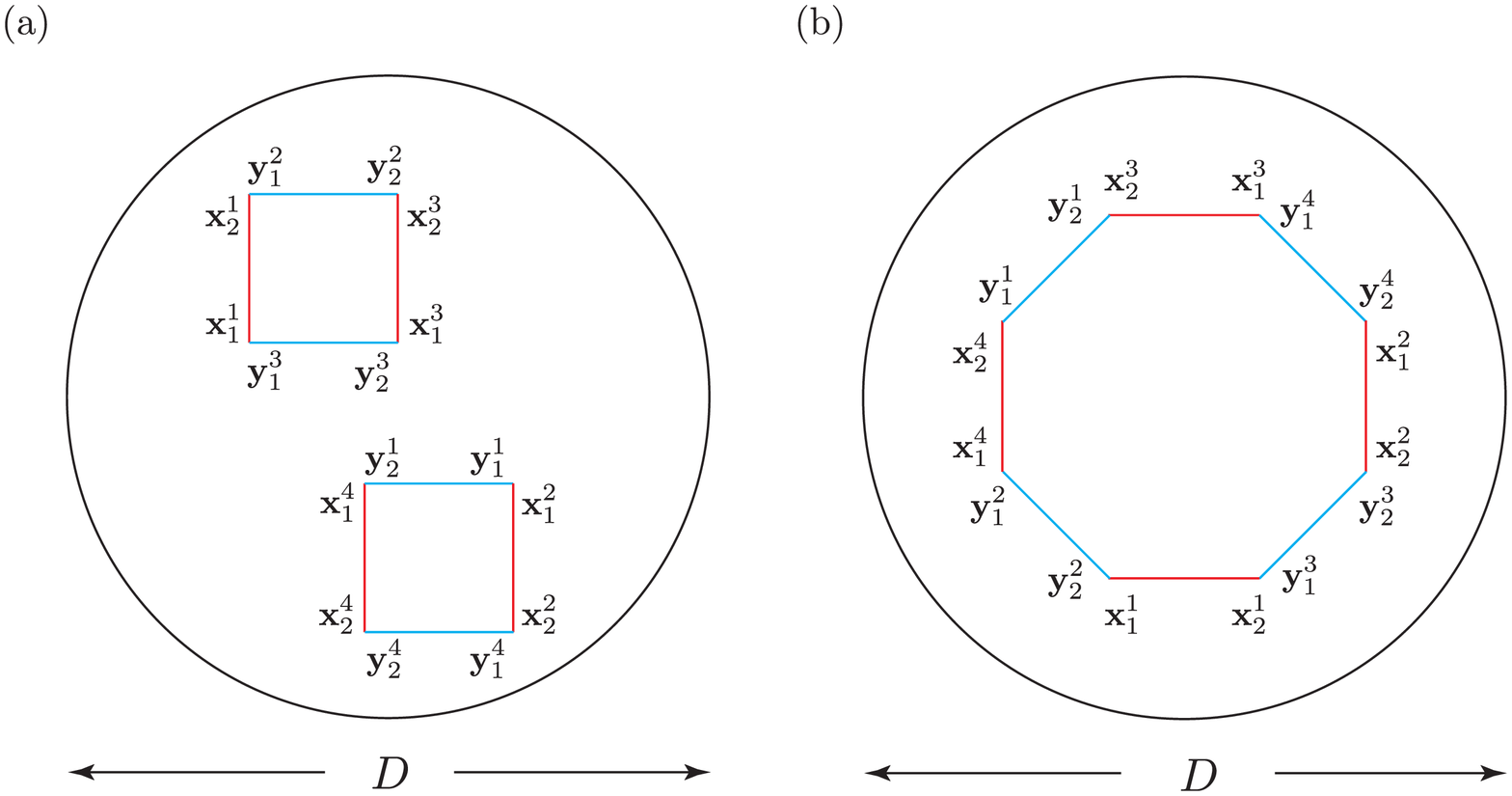}
\caption{Two possible configurations for a non-zero matrix element $\langle(\hat{P}^\dagger)^N\hat{P}^N\rangle$.}
\label{fig:element}
\end{figure}

To understand the physics of the mean-field problem, we use a variational wave function, motivated by the BCS theory
\cite{BCS}:
\begin{equation}
\label{6}
|\Psi_{\rm BCS}\rangle
=\sum_N  \frac{\tilde\Delta^N \hat P^N}{(2N)!}|0\rangle.
\end{equation}
where $\tilde\Delta$ is a complex order parameter, and $|0\rangle$ is the vacuum state with no bosons. We now find the normalization of the wave function (\ref{6}):
\begin{equation}
\label{7}
\langle\Psi_{\rm BCS}|\Psi_{\rm BCS}\rangle=\sum_N |\tilde \Delta|^{2N}\frac{\langle (\hat P^\dagger)^N\hat  P^N\rangle }{[(2N)!]^2}.
\end{equation}
The matrix element $\langle (\hat P^\dagger)^N \hat P^N\rangle=\int\Pi_{i=1}^N d^2x^{i}_1d^2x^{i}_2d^2y^i_1d^2y^i_2\langle \Pi_i \hat p^\dagger({\bf x}^i_1,{\bf x}^i_2)\Pi_i\hat p({\bf y}^i_1,{\bf y}^i_2) \rangle$ is represented in Fig.~\ref{fig:element}. $N$ operators $\hat p({\bf y}^i_1,{\bf y}^i_2)$ are shown with blue lines connecting ${\bf y}^i_1$ with ${\bf y}^i_2$. Red lines from ${\bf x}^i_1$ to ${\bf x}^i_2$ depict the operators $\hat p^\dagger({\bf x}^i_1,{\bf x}^i_2)$. Non-zero average requires that after reordering the indexes $i$, ${\bf x}^i_{1,2}$ become approximately equal to ${\bf y}^j_{1,2}$. This can be arranged in $(2N)!$ ways. The loops in Fig.~\ref{fig:element} can be understood as diagrams in algebraic theory of anyons. Since $B$ has a trivial topological spin, a trivial Frobenius-Schur indicator, and the quantum dimension of 1, the topological factor, associated with each diagram, is 1. We assume such choice of the operator $\hat P$ that no Abelian Aharonov-Bohm phases are associated with closed loops.
A non-topological factor comes from the overlap integrals of the wave functions, generated by $B$ at nearby points (those wave functions are not orthogonal). Thus, the diagrams carry real factors $a^{2N}$. After defining $\Delta=a\tilde\Delta$, we find
\begin{equation} \label{8}
\langle\Psi_{\rm BCS}|\Psi_{\rm BCS}\rangle=\cosh|\Delta|.
\end{equation}

To find the energy, we need to compute the averages of $\hat P$, $\hat P^\dagger\hat P$, and $N_B$. For $N_B$, we find
\begin{align} \label{9}
\nonumber
N_B
&=~\frac{1}{\langle\Psi_{\rm BCS}|\Psi_{\rm BCS}\rangle}\sum_N 2N |\tilde \Delta|^{2N}\frac{\langle (\hat P^\dagger)^N\hat  P^N\rangle }{[(2N)!]^2}
\\
&=~|\Delta|\tanh|\Delta|.
\end{align}
We are interested in the limit of large $N_B$, so that $N_B\approx|\Delta|$, $\tanh\Delta\approx 1$, $\cosh|\Delta|\gg 1$. In that limit we easily find the averages of $\hat P$ and $\hat P^\dagger\hat P$:
\begin{equation}
\label{10}
\frac{\langle\Psi_{\rm BCS}|\hat P|\Psi_{\rm BCS}\rangle}{\langle\Psi_{\rm BCS}|\Psi_{\rm BCS}\rangle}=a\Delta^*,
\end{equation}

\begin{equation}
\label{11}
\frac{\langle\Psi_{\rm BCS}|\hat P^\dagger \hat P|\Psi_{\rm BCS}\rangle}{\langle\Psi_{\rm BCS}|\Psi_{\rm BCS}\rangle}\approx a^2|\Delta|^2.
\end{equation}
Redefining the coefficients of the Hamiltonian $U=a\tilde U$ and $W=a^2\tilde W$, introducing the boson density $n=N_B/S$, and rewriting $\Delta=N_B\exp(i\theta)$, we find the average energy
\begin{equation}
\label{12}
E=S[Un\cos\theta+Wn^2+n\epsilon_0/2]. 
\end{equation}
For $U\gg\epsilon_0$, the ground state is a Bose-Einstein condensate with $\theta=\pi$ and $n=U/2W$.

\end{document}